\documentclass{iopart}
\usepackage{iopams}

\newcommand{\ndims}{\mbox{($4\!+\!n$)\,}}
\newcommand{\be}{\begin{equation}}
\newcommand{\ee}{\end{equation}}
\newcommand{\da}{\dot{a}}
\newcommand{\db}{\dot{b}}

\newcommand{\dda}{\ddot{a}}
\newcommand{\ddb}{\ddot{b}}

\newcommand{\fda}{\frac{\da}{a}}
\newcommand{\fdb}{\frac{\db}{b}}

\newcommand{\fdda}{\frac{\dda}{a}}
\newcommand{\fddb}{\frac{\ddb}{b}}

\newcommand{\fkaa}{\frac{k_a}{a^2}}
\newcommand{\fkbb}{\frac{k_b}{b^2}}

\begin{document}
\title{Can homogeneous extra dimensions be stabilized during 
       matter domination?}
\author{Torsten Bringmann and Martin Eriksson}
\address{Department of Physics, AlbaNova University Center, Stockholm
       University, SE - 106 91 Stockholm, Sweden}
\eads{\mailto{Torsten.Bringmann@physto.se}, \mailto{mate@physto.se}}
\date{August 21, 2003}
\begin{abstract}
We investigate cosmologies with homogeneous extra dimensions that can be described by generalized Friedmann-Robertson-Walker metrics and give a brief review on a general setup to describe a broad range of standard stabilization and compactification mechanisms. These mechanisms allow for solutions to the field equations with static extra dimensions if the universe is dominated by radiation or an approximately constant energy-momentum tensor, as for example due to a cosmological constant. During matter domination, however, there are no static solutions for the extra dimensions and we therefore conclude that in this setup it is not possible to construct stabilization mechanisms for the \emph{whole} evolution history of the universe. We furthermore show in detail that the two possible approaches of discussing either the higher-dimensional field equations directly or the dimensionally reduced and conformally transformed theory, are exactly equivalent and thus lead to the same conclusions. Finally, we indicate a possible way out of these difficulties.
\end{abstract}
\maketitle

\section{Introduction}

The idea that our world may consist of more than four space-time dimensions, with extra spatial dimensions compactified on some small scale, goes back to Nordstr\"om, Kaluza and Klein \cite{nor}. One of the main motivations for these and subsequent works was the hope for a possible unification of all interactions (see e.g.~\cite{bai87} for a nice review on Kaluza-Klein theories). Recent years have seen a great revival of interest in extra-dimensional scenarios, most notably due to the influence of string theory. For a review on different models with extra dimensions see for example \cite{rub} and references therein.

An important testing ground for theories involving extra dimensions is cosmology. A generic prediction of these theories is that (some of) the fundamental coupling constants vary with the volume of the internal space. However, the strong cosmological constraints on the allowed variation of these 'constants' (see for example \cite{oli}) require the extra space to be not only compactified, but also stabilized, at a time no later than big bang nucleosynthesis (BBN). One of the main tasks is therefore to find a dynamical explanation for this.  Of course, one should also be able to reproduce standard cosmology for late times. An interesting idea is furthermore the possibility of a dark matter candidate arising from such theories \cite{sla,cheng}.

Recently, the stability properties of so-called universal extra
dimensions \cite{appa} have been investigated in the absence of any
explicit stabilization mechanism. It was shown that while static extra
dimensions arise naturally during radiation domination, for matter
domination there are no solutions that reproduce standard cosmology
\cite{bri}. Here, we will consider a much more general setup,
including not only the case of universal extra dimensions, and take
into account a wide range of available stabilization mechanisms for
homogeneous extra dimensions. We find that the problems with a matter
dominated universe as encountered before seem to be generic: none of
the mechanisms considered can account for stable extra dimensions and
a standard cosmological evolution of ordinary space during an era of
matter domination. For previous work on the dynamical evolution and
stability properties of homogenous extra dimensions see
e.g.~\cite{frea,cho,randj,mae,gun,car,gun2} and references therein. 

This paper is organized as follows: In Section \ref{cosmo} we give a
brief introduction to cosmology with homogeneous extra dimensions and how to get the Friedmann-like equations that describe the evolution of the higher-dimensional metric. Section \ref{stable} reviews the general form of the stabilization mechanisms considered, including concrete examples, and Section \ref{standard} shows that while standard cosmology during radiation domination is thereby achieved easily, none of these mechanisms work for matter domination. An alternative way to investigate higher-dimensional theories is to dimensionally reduce the action and then to perfom a conformal transformation in order to recover (in the present context) ordinary four-dimensional general relativity plus a scalar field. In Section \ref{dimred} we review this procedure and show explicitly that these two approaches are equivalent and lead to the same predictions. Finally, Section \ref{conc} concludes.

\section{Cosmology with homogeneous extra dimensions}
\label{cosmo}
In the following we shall consider a \ndims-\,dimensional spacetime of the form $F^4\times K$, with $F^4$ the observed Friedmann solution and $K$ a compact $n$-dimensional manifold. Our sign conventions are those of \cite{mis}. We adopt coordinates $X^A$, $A = 0, 1, \ldots, 3 + n$ with
\be
  x^\mu \equiv X^\mu \quad (\mu = 0, 1, 2, 3) \quad\mbox{and}
  \quad x^i \equiv X^i \quad (i = 1, 2, 3) 
\ee
being the coordinates for ordinary four-dimensional spacetime and
three-dimensional space (3D) respectively, and
\begin{equation}
  y^p \equiv X^{3+p} \quad (p = 1, \ldots, n) \,,
\end{equation}
the coordinates for the extra dimensions. Einstein's field equations are given by
\begin{equation}
  R_{AB} - \frac{1}{2} R\, g_{AB}+\Lambda\,g_{AB}=
  \kappa^2 T_{AB}\,, \label{fe} 
\end{equation}
where the  coupling constant $\kappa^2$ is related to its four-dimensional analogue $\bar\kappa^2$ via the $n$-dimensional volume $\tilde V$ of the compact space $K$:
\be
  \label{kappa} 
  \kappa^2 = \tilde V\, \bar\kappa^2\,.
\ee 

To describe the evolution of the higher-dimensional universe it is natural to assume a metric with \emph{two} time-dependent scale factors $a(t)$ and $b(t)$ \cite{bai87}:
\begin{equation}
  g_{AB}~\rmd X^A\rmd X^B = - \rmd t^2 + a^2(t) \gamma_{ij}~\rmd x^i
    \rmd x^j + b^2(t) \tilde{\gamma}_{pq}~\rmd y^p \rmd y^q \,,
    \label{metric}
\end{equation}
where $\gamma_{ij}$ and $\tilde{\gamma}_{pq}$ are maximally symmetric
metrics in three and $n$ dimensions, respectively. Spatial curvature
is thus parametrized in the usual way by $k_a = -1,0,1$ in ordinary
and $k_b = -1,0,1$ in the compactified space. This choice of metric is
certainly not the most general one but it clearly illustrates the
difficulties of stabilization during matter domination. Taking the
compact space to be an Einstein space does not change our results. For
a thorough analysis of models with anisotropic homogeneous extra
dimensions, see \cite{dem} and references therein.

The choice of the metric determines the form of the energy-momentum tensor to be
\be
\label{emtensor} 
  T_{00} = \rho\,, \qquad T_{ij}=-p_a\gamma_{ij}\,,\qquad
  T_{3+p\,3+q}=-p_b\tilde\gamma_{pq}\,,
\end{equation}
which describes a homogeneous but in general anisotropic perfect fluid
in its rest frame. The non-zero components of the field equations~(\ref{fe}) are then given by\footnote{
These expressions become equivalent to those in \cite{frea} via the trivial rescaling $k_b\rightarrow\frac{2}{n-1}k_b$.}
\begin{eqnarray}
  \fl 3 \left[ \left( \fda \right)^2 + \fkaa \right] + 3n \fda \fdb +
  \frac{n(n - 1)}{2}\left[\left( \fdb \right)^2 + \fkbb \right] =
  \Lambda + \kappa^2 \rho \,, \label{fe00} \\
  \fl 2\fdda + \left( \fda \right)^2 + \fkaa + 2n \fda
  \fdb + n \fddb + \frac{n(n - 1)}{2} \left[ \left( \fdb\right)^2 +
  \fkbb \right] = \Lambda - \kappa^2 p_a \,, \label{feij} \\
  \fl 3\fdda + 3 \left[ \left( \fda \right)^2 + \fkaa \right] +3(n -
  1) \fda \fdb + (n - 1) \fddb + \frac{(n - 1)(n -2)}{2}\left[ \left(
  \fdb \right)^2 + \fkbb \right] \nonumber \\
  = \Lambda - \kappa^2 p_b \,,
  \label{fepq}
\end{eqnarray}
where a dot denotes differentiation with respect to cosmic time
$t$. For later convenience, we will not use (\ref{fepq}) but instead
\be
  \label{newf3}
  \fddb + 3\fda\fdb + (n-1)\left[\left(\fdb \right)^2+\fkbb\right] 
  = \frac{2\Lambda}{n+2} + \frac{\kappa^2}{n+2}\left(\rho-3p_a+2p_b\right)\,,
\ee
which is a linear combination of equations~(\ref{fe00}\,-\,\ref{fepq}).
 
The energy density can be related to the pressures by the equations of
state $p_a=w_a\rho$ and $p_b=w_b\rho$. From conservation of energy
$T^A_{\phantom{A}0;A} = 0$ it then follows that
\begin{equation}
  \frac{\dot{\rho}}{\rho} = -3(1 + w_a) \fda - n(1 + w_b) \fdb\,.
  \label{econserv}
\end{equation}
For constant $w_a$, $w_b$ this can be integrated to
\begin{equation}
  \rho = \rho_0 a^{-3(1 + w_a)} b^{-n(1 + w_b)}\,. \label{rho} 
\end{equation}

\section{Compactification and stabilization of extra dimensions}
\label{stable}

We do not directly observe any extra dimensions, so we expect the space $K$ to be compactified on some small scale. (For a non-compactified approach see, e.g. \cite{ran}.) Furthermore, from (\ref{kappa}), any time-variation of $b$ would lead to a time-dependent (four-dimensional) gravitational coupling 'constant'. As there are tight constraints on such a behaviour, the size of the compact space should be stabilized already at an early stage of the cosmological evolution, i.e.~no later than BBN. One usually assumes that both of these requirements can be attained dynamically by introducing background fields. Since their role is to separate ordinary space from the extra dimensions they typically contribute an effective action of the form
\be
  \label{sbg}
  S^{\mathrm{bg}}=-\int\rmd^{4+n}X\sqrt{-g}\,W(b)
\ee
to the theory \cite{bai84a,bai87}. Even in the simple case of
\be
  \label{powerw}
  W(b)\propto b^{-m}\,,
\ee
one may thereby describe the stabilizing effect of gauge-fields
wrapped around two extra dimensions ($m=4$) \cite{cre} and the
generalization of this to the Freund-Rubin mechanism ($m=2n$)
\cite{freb}, or Casimir energy of massless fields ($m=n+4$) \cite{appb,pon}. Examples that can not be described by the simple form (\ref{powerw}) but still fit into the general scheme (\ref{sbg}) include for example the Casimir energy of massive fields in orbifold theories \cite{pon} or scenarios where several different mechanisms contribute to $S^{\mathrm{bg}}$.

The energy-momentum tensor corresponding to the action (\ref{sbg}) can be computed from
\be
  \delta S^{\mathrm{bg}}=\frac{1}{2}\int\rmd^{4+n}X
  \sqrt{-g}\,T^{\mathrm{bg}}_{AB}\delta g^{AB}
\ee
and takes the form
\be
\label{tbg} 
  T^{\mathrm{bg}}_{00} = \rho^{\mathrm{bg}}\,, 
  \qquad T^{\mathrm{bg}}_{ij}=-p^{\mathrm{bg}}_a\gamma_{ij}\,,
  \qquad T^{\mathrm{bg}}_{3+p\,3+q}=-p^{\mathrm{bg}}_b\tilde\gamma_{pq}\,,
\ee
with
\begin{equation}
\label{rpbg1}
    \rho^{\mathrm{bg}} = -p^{\mathrm{bg}}_a = W(b)
\end{equation}
  and
\begin{equation} 
\label{rpbg2}
    -p^{\mathrm{bg}}_b = W(b) + \frac{b}{n}W'(b)\,.
\end{equation}

\section{Recovering standard cosmology}
\label{standard}

Let us assume that a stabilization mechanism due to background fields
is at work, i.e.~$b(t)=b_0$ is constant at late times. Since these
fields also contribute to the energy-momentum tensor, we replace
$\rho\rightarrow\rho+\rho^{\mathrm{bg}}$ and $p_{a,b}\rightarrow
p_{a,b}+p_{a,b}^{\mathrm{bg}}$ in the field equations
(\ref{fe00}\,-\,\ref{fepq}). All contributions to the energy-momentum
tensor that are not due to background fields or a cosmological
constant are then given by $\rho$ and $p_{a,b}$, and equations
(\ref{fe00}\,-\,\ref{feij}) reduce to the standard Friedmann equations
with an effective four-dimensional cosmological constant
\cite{randj,bai87} 
\be
  \label{lambda4}
  \bar\Lambda = \frac{2\Lambda}{n+2} + \frac{\kappa^2}{n+2}\,
     \Big[2W(b_0)+b_0W'(b_0)\Big].
\ee
(Note that $\kappa^2\rho=\bar\kappa^2\bar\rho$, where  $\bar\rho$ is
     the four-dimensional energy density; in the same way
     $\kappa^2p_a=\bar\kappa^2\bar p$.)  It now just remains to check,
     whether the remaining equation (\ref{newf3}) is consistent:
\be
  \label{f3const}
  \fl(n+2)(n-1)\frac{k_b}{b_0^2} = 2\Lambda + 
  2\kappa^2\left[W(b_0)-\frac{b_0}{n}W'(b_0)\right]+
  \kappa^2\left(\rho-3p_a+2p_b\right)\,.
\ee

For an era where the energy momentum tensor is dominated by usual
four-dimensional radiation, the last term vanishes since $\rho=3p_a$
and $p_b=0$. This should be the case after compactification and
stabilization has taken place and we have $a\gg1/T\gg b$. Then, for
any given $W$ and $b_0$, equation (\ref{f3const}) can be satisfied by
fine-tuning the value of the higher-dimensional cosmological constant
$\Lambda$. Requiring furthermore that the effective four-dimensional
cosmological constant $\bar\Lambda$ given by (\ref{lambda4}) is
negligible during radiation domination, corresponds to $b_0$ being a
stationary point of
\be
  W_\mathrm{eff}(b)\equiv W(b)-
  \frac{n(n-1)}{2\kappa^2}\frac{k_b}{b^2}\,,
\ee
i.e.~$W'_\mathrm{eff}(b_0)=0$.
In that way standard cosmological evolution is reproduced during radiation domination, without a cosmological constant. If $b_0$ is a local \emph{minimum} of $W_\mathrm{eff}(b)$ the solution is also stable against small perturbations \cite{bai84a}.

However, \eref{f3const} can only be satisfied if \footnote{
This constraint also appears in \cite{ell,kan,bri}. Here, however, it applies to the components of the energy-momentum tensor \emph{after} the contributions from the stabilization mechanism have been subtracted.}
\be
  \label{eof}
  \rho - 3p_a + 2p_b = const,
\ee
and it is hard to see how this can possibly be the case if the energy-momentum tensor is not dominated by four-dimensional radiation.
For example, throughout most of its evolution the universe has been dominated by non-relativistic matter with negligible pressure $p_a\ll\rho$. The only way to get static extra dimensions during such an epoch would be to have 
\be
  p_b=-\frac{1}{2}\rho+const.
\ee
Since $\rho$ is the \emph{total} energy density, such an equation of state seems highly contrived and we will not consider it further. A hint at what it might correspond to is given by \cite{gu}.

Of course, condition (\ref{eof}) is obviously satisfied if the energy density and the pressures are constant in time. Recent type Ia supernovae observations \cite{ton} strongly suggest that the presently dominating energy component has negative pressure and this may well be due to a cosmological constant, which is characterized by constant energy density and pressure. However, such a vacuum energy is negligible up until relatively recently and can thus not solve the problems with a long period of matter domination.

\section{Dimensional reduction}
\label{dimred}
So far, we have directly discussed the field equations (\ref{fe}). A complementary way of studying multi-dimensional theories is to dimensionally reduce the action by integrating out the internal dimensions and then perform a Weyl scaling. To be more precise, one starts from the action,
\be
  \label{action1}
  S=\frac{1}{2\kappa^2}\int\rmd^{4+n}X\sqrt{-g}
    \left(R - 2\Lambda - 2\kappa^2\mathcal{L}_\mathrm{matter}\right)\,,
\ee
and assumes a metric of a slightly more general kind than (\ref{metric}):
\be
 g_{AB}~\rmd X^A \rmd X^B = \bar g_{\mu\nu}~\rmd x^\mu \rmd x^\nu
    +b^2(x^\mu)\tilde g_{pq}~\rmd y^p \rmd y^q \,, 
\ee
where $\tilde g_{pq}$ depends on the internal coordinates $y^p$ only. Decomposing the curvature scalar $R$, integrating out the extra dimensions and discarding all total divergences gives
\be
  \label{action2}
  \fl S=\frac{\tilde{\mathcal{V}}}{2\kappa^2}\int\rmd^{4}x\sqrt{-\bar g}~b^n
    \left(\bar R +b^{-2}\tilde R + n(n-1)b^{-2}\partial_\mu b \partial^\mu b 
    - 2\Lambda - 2\kappa^2\mathcal{L}_\mathrm{matter}\right)\,,
\ee
where $\tilde{\mathcal{V}}=\int\rmd^{n}y\sqrt{\tilde g}$ while $\bar R$ and $\tilde R$ are the curvature scalars constructed from $\bar g_{\mu\nu}$ and $\tilde g_{pq}$, respectively. For $\tilde g_{pq}=\tilde \gamma_{pq}$ one gets $\tilde R=n(n-1)k_b$.

The next step is to put the gravitational part of this action in standard Einstein-Hilbert form by performing a Weyl transformation to a new metric $\hat g_{\mu\nu}=b^n \bar g_{\mu\nu}$. Introducing $\hat\kappa^2=\kappa^2/\tilde{\mathcal{V}}$ and $\Phi=\sqrt{\frac{n(n+2)}{2\hat\kappa^2} }\ln{b}$ then results in four-dimensional gravity plus a scalar field,
\be
  \label{action3}
  S=\int\rmd^{4}x\sqrt{-\hat g}
    \left(\frac{1}{2\hat\kappa^2} \hat R 
    - \frac{1}{2}\partial_\mu\Phi \partial^\mu\Phi + V_{\mathrm{eff}}(\Phi)\right)\,,
\ee
with an effective potential
\begin{eqnarray}
  \fl V_{\mathrm{eff}}(\Phi)=\frac{n(n-1)k_b}{2\hat\kappa^2}~
  \exp \left( -\sqrt\frac{2(n+2)\hat\kappa^2}{n}\Phi \right) \nonumber \\
  -\left(\frac{\Lambda}{\hat\kappa^2}+\tilde{\mathcal{V}}
  \mathcal{L}_\mathrm{matter}\right) \exp
  \left(-\sqrt\frac{2n\hat\kappa^2}{n+2}\Phi\right)\,.  
\end{eqnarray}
The fact that the four-dimensional gravitational coupling constant
$\hat\kappa^2$ above differs from the one in (\ref{kappa}) and
(\ref{action2}) is an artifact of the conformal transformation. 

The theories described by (\ref{action1}) and (\ref{action3}) are
completely equivalent, i.e.~they result in the same equations of
motion. For the scalar field in (\ref{action3}), for example, one has
\be
  \label{box}
  \hat\Box\Phi=-V_{\mathrm{eff}}'(\Phi)\,,
\ee
where $\hat\Box=\frac{1}{\sqrt{-\hat g}}\partial_\mu\sqrt{-\hat
  g}\partial^\mu$ is the Laplace-Beltrami operator for the metric
$\hat g_{\mu\nu}$. In order to compare this with the results of the
previous sections, we restrict ourselves to the metric (\ref{metric}) and write the higher-dimensional matter part in perfect fluid form $\mathcal{L}_\mathrm{matter}=\rho=\rho_0a^{-3(1+w_a)}b^{-n(1+w_b)}$. One may now verify that (\ref{box}) gives exactly (\ref{newf3}),
from which all previous results follow.\footnote{
It is, however, important to keep in mind that after the conformal transformation the independent variables of $\mathcal{L}_\mathrm{matter}$ are no longer $a$ and $b$ but $\hat a=b^\frac{n}{2} a$ and $b$ (appropriately expressed in $\Phi$). The $b$-dependence of $\rho$ thus changes from $\rho\propto b^{-n(1+w_b)}$ to $\rho\propto b^{\frac{n}{2}(1+3w_a-2w_b)}$.
}

The equation of motion (\ref{box}) for the scalar field thus provides a convenient way to look for static solutions of the scale factor $b$. Since the spatial derivatives vanish, such solutions must obey 
\be
  V_\mathrm{eff}'(\Phi)=0\,,  
\ee
and if the extremal point of $V_\mathrm{eff}(\Phi)$ is a minimum, one furthermore expects these solutions to be stable. 
However, caution must be taken as to which metric one regards as physical, i.e.~whether $\bar g_{\mu\nu}$ or $\hat g_{\mu\nu}$ is taken to have FRW-form and thus how the matter part couples to the scale factors. Another important point to notice is that $V_\mathrm{eff}(\Phi)$ is time-dependent through its dependence on $\mathcal{L}_\mathrm{matter}$. So even if there existed a minimum for the effective potential at all times of the cosmological evolution, its value would still be expected to vary during transition periods such as from radiation to matter domination. The scale factor $b$ would then no longer stay constant but  be driven towards the new minimum according to (\ref{box}).

\section{Conclusions}
\label{conc}
 
 One of the main quests of higher-dimensional cosmology is to
 understand the special role of the extra dimensions compared to the
 evolution of ordinary three-dimensional space. Strong observational
 constraints on the variation of coupling constants require, in
 particular, that the extra dimensions are effectively stabilized no
 later than BBN. Such a stabilization is usually achieved by
 introducing background fields, typically the same fields that are
 supposed to be responsible for the compactification of the internal
 space. A wide range of these  mechanisms can be expressed in a
 phenomenological way by adding a term of the form (\ref{sbg}) to the
 action. This term only depends on the size of the extra dimensions,
 which is what one would expect since its role is to separate ordinary
 three-space from the internal space. Concrete examples of mechanisms
 fitting into this scheme include those described in
 \cite{cre,freb,appb,pon}. 

By analysing the higher-dimensional field equations or, equivalently, the equations of motion for the dimensionally reduced theory, it can be seen that this general setup provides the possibility of having static extra dimensions and a standard cosmological evolution of ordinary three-space -- at the cost of fine-tuning the higher-dimensional cosmological constant. However, this result holds only if the universe is either dominated by radiation or a constant energy-momentum tensor. For a matter-dominated universe no static solutions are available and thus this setup cannot be used to explain the stability of the extra dimensions during the whole evolution of the universe.

The only possibility to get homogeneous static extra dimensions compatible with standard cosmology is therefore to invent stabilization mechanisms that -- in contrast to (\ref{sbg}) -- not only depend on the size of the extra dimensions but, for example, also on the total energy density $\rho$. Indeed, allowing for interaction terms between the background fields and the matter Langrangian, that is precisely what one would expect. Such a dependence on a second time-varying quantity would introduce additional time-dependent terms in the analogue of (\ref{f3const}) and thus in principle allow for static solutions for a suitably chosen $W(b,\rho)$. In such a scenario, however, the  minima of the effective potential $V_\mathrm{eff}$ are still expected to depend on the dominant contribution to the energy density and the size of the extra dimensions should therefore change during transition periods like for example from radiation to matter domination. 

Finally, we would like to stress that this article considers homogeneous cosmologies. In brane world scenarios for example, there may be other stabilization mechanisms at hand that naturally fulfil the constraint (\ref{eof}) -- see for example \cite{kan} for a corresponding suggestion. 
For the reasons given above, however, a variation of the size of the extra dimensions should be expected for \emph{all} models that allow for changing equations of state in the bulk.

\ack
We are grateful to Michael Gustafsson, Lars Bergstr\"om and
Stefan Hofmann for helpful discussions and careful reading of the
manuscript.


\end{document}